\documentclass[12pt]{article}
\usepackage{amssymb,amsmath}
\begin{document}
\begin{flushright}
\end{flushright}
\vspace{10mm}
\begin{center}

{\Large \bf Asymptotic representations for some functions and integrals
connected with the Airy function}\\

 \vspace{10mm} A.I. Nikishov and V.I. Ritus ${}^\dag$\\

\vspace{3mm}{$\dag$\it Lebedev Physical Institute, 119991, Moscow, Russia\\
             e-mail: nikishov@lpi.ru; ritus@lpi.ru}

    \vspace{2mm}

    \end{center}
        \begin{abstract}
        The asymptotic representations of the functions ${\rm Ai}_1(x)$,
        ${\rm Gi}(x)$, ${\rm Gi}'(x), {\rm Ai}^2(x)$, ${\rm Bi}^2(x)$ are 
        obtained. As a by-product, the factorial identity ($21'$) is found.
        The derivation of  asymptotic  representations of the integral
        $\int_v^{\infty}dx{\rm Ai}(x)h(x,v)$ for $v\to-\infty$ and integrals,
        differing from it by the change of ${\rm Ai}(x)$ by ${\rm Ai}'(x)$
        or ${\rm Ai_1(x)}$, is presented. For the Airy function ${\rm Ai}(z)$,
        as an example, the Stokes' phenomenon is  considered as a consequence
        of discontinuous behavior of steepest descent lines over the passes.
        When $z$ crosses the Stokes ray, the steepest descent line over the
        higher pass abruptly changes the direction of its asymptotic approach
        to the steepest descent line over the lower pass to the direction of
        approach to the opposite end of this line. Therefore, when the
        integration contour, drawn along the steepest descent lines, goes over
        the higher pass, it begins or stops to go over the lower pass while 
        $z$ crosses the Stokes ray, and as a result the recessive series
        (contribution from the lower pass) discontinuously appears or 
        disappears in the asymptotic representation of a function containing
        the dominant series.
         
        \end{abstract}

        \section{Introduction}

 In the theory of quantum processes with particles in a constant
 or slowly varying field, the Airy function, related functions
 and integrals with these functions play an important role [1, 2].
 In this paper [which is the Lebedev Phys. Inst. Preprint N 253 (1985)]
  we give the asymptotic expansions for some frequently
  occurring functions. 

 \section{Asymptotic behavior of ${\rm Hi}(z)$, ${\rm Gi}(z)$, ${\rm Ai}_1(z)$
  and derivatives ${\rm Hi}'(z)$, ${\rm Gi}'(z)$}

  We define the Airy function and related functions as in [3, 4], 
  but without the factor $\pi^{-1}$. Starting from 
  $$
   {\rm Hi}(z)=\int_0^{\infty}dt\,e^{zt-\frac{t^3}3}  \eqno(1)
   $$
   and integrating by parts, we have
   $$
    {\rm Hi}(z)=\sum_{k=1}^nq^{(k-1)}(0)(-z)^{-k}+\varepsilon_n(z), \quad
    \varepsilon_n(z)=(-z)^{-n}\int_0^{\infty}dt\,q^{(n)}(t)e^{zt},  \eqno(2)
   $$
   $$
    q(t)=e^{-t^3/3},\quad q'(t)=-t^2q(t),
    $$
    $$
    q''(t)=(-2t+t^4)q(t),\quad q^{(3)}(t)=(-2+6t^3-t^6)q(t),\,\ldots\,. \eqno(3)
    $$

  The relation (2) holds for any $z$, but it is useful only
  for ${\rm Re}\, z\ll-1$ (more exactly for $|z|\to\infty$, 
  $|{\rm ph}(-z)|<2\pi/3$, see p.\,432 in [4], when the remainder term 
  $\varepsilon_n(z)$ is small compared with the terms of the sum).
 This is not the case for $z\gg1$ in (2). So the decrease of
  terms in the sum (2) does not provide any guarantee of smallness 
  of the remainder term [4, 5]. Its magnitude depends on proximity 
  of saddle points (in general, critical points) to the integration
  path. Assuming $z$ fixed and ${\rm Re}\, z\ll-1$ we shall
  increase $n$ in (2). The terms of the sum first decrease in 
  magnitude then increase, as the representation is asymptotic.
  It is clear that the sum of increasing terms, generally speaking,
  grows with $n$ and this must be compensated by the remainder term;
  the representation with such a term is still exact. When 
   $n\to\infty$ and the remainder term is dropped, we get the
   asymptotic series and the problem of interpretation of
   diverging part of the series arises [5]. Dingle [5] suggested 
   that the remainder term can be restored just from the general 
   term in each considered series. Indeed, the remainder term 
   can be restored using, when necessary, the additional information
   about the function. For this reason it is convenient to regard
   the divergent sum of late terms as a symbolic representation
   of the remainder term [5]. With this understanding the
   asymptotic series uniquely represents its function and is
   the complete asymptotic expansion (for a certain phase
   range of $z$). In general, the complete asymptotic expansion 
   consist of several series (two for solutions to second-order
   homogeneous differential equations) each of which determines 
   its own function [5]. In this case the complete asymptotic
   expansion expresses the linear dependence of the initial
   function upon other functions each of which is represented
   by its own series in the considered sector of $z$.
   This linear dependence between the functions holds in all
   sectors of the complex variable $z$, but outside the
   considered sector the representation of each function by
   one series may be unsatisfactory, that is the remainder  term 
   may be not small.

   Asymptotic expansions with remainder terms exactly determine
   their functions. However, even in the case when the remainder
   terms are represented only symbolically by the increasing
   terms of infinite series, it is still possible to require,
   that complete asymptotic expansions satisfy the same relations
   as the functions themselves [5].

   Returning now to eq.(2), for $n\to\infty$ we have [4]
   $$
    {\rm Hi}(z)\sim-\sum_{n=0}^{\infty}\frac{(3n)!}{3^nn!z^{3n+1}},\quad
     |{\rm ph}(-z)|<\frac{2\pi}3.                               \eqno(4)
   $$
                 
  Another and more simple way of obtaining (4) consists of 
   expanding $\exp(-t^3/3)$ in power series and subsequent 
   term by term integration. The integrable series is convergent
   everywhere, but nonuniformly , so the integrated series is asymptotic.

   We consider now the real $z=x\gg1$. The integrand in (1) 
   first exponentially increases (up to the saddle point
   $t_c=\sqrt x$) then decreases. In accordance with this the 
    asymptotic expansion consists of a series of exponentially 
    large terms (contribution from the saddle point) and, as
    we shall see below, of the series (4) (contribution from 
    the lower limit of integration).

     As mentioned above, complete
     asymptotic expansion should satisfy the same relations as the
    functions themselves. The relation
    $$
   {\rm Ai}_1(x)=\pi+{\rm Hi}(x){\rm Ai}'(x)-{\rm Hi}'(x){\rm Ai}(x),\eqno(5)
     $$
     where
      $$
      {\rm Ai}_1(x)=\int_x^{\infty}dt\,{\rm Ai}(t), \quad
       {\rm Ai}(x)=\int_0^{\infty}dt\,\cos\left( xt+\frac{t^3}3\right),
      \eqno(6)
       $$
       is important for us here. It is easily verified by 
       differentiation and using the equations
   $$
     {\rm Hi}''(x)-x{\rm Hi}(x)=1, \quad {\rm Ai}''(x)-x{\rm Ai}(x)=0. \eqno(7)
    $$

 We shall obtain the asymptotic expansions for ${\rm Hi}(x)$, 
  ${\rm Hi}'(x)$ and  ${\rm Ai}_1(x)$ and check them up using (5) and
    the known asymptotic expansions for ${\rm Ai}(x)$ and
      ${\rm Ai}'(x)$ [3, 4].

       The complete asymptotic expansion for ${\rm Hi}(x)$, $x\to\infty$,
       can be obtained in the following manner. Substituting
       $t=\tau+\sqrt x$, we have for the exponent in (1)
      $$
       xt-\frac{t^3}3=\frac23x^{3/2}-x^{1/2}\tau^2-\frac{\tau^3}3,
        \quad \tau=t-\sqrt x.                                   \eqno(8)
         $$
        Next we expand $\exp(-\tau^3/3)$ in power series and 
       integrate term by term over $\tau$ from $-\sqrt x$ to
        $\infty$. If we simply extend the region of integration 
        to the whole real axis (using the condition $\sqrt x\gg1$),
        we get only the exponentially large terms and the relation
        (5) will not be satisfied. So we write
       $$
        {\rm Hi}(x)=\int_{-T}^{\infty}dt\,e^{xt-\frac{t^3}3}-
        \int_{-T}^0dt\,e^{xt-\frac{t^3}3},\quad T\gg1.              \eqno(9)
        $$
     In the first integral we use (8) and perform the indicated operations.
In the second integral we expand $\exp(-\tau^3/3)$ in power series and
integrate  term by term. Letting $T\to\infty$, we get
$$
{\rm Hi}(x)\sim\pi^{1/2}x^{-1/4}e^{\zeta}\sum_{n=0}^{\infty}\frac{c_n}{\zeta^n}-
\sum_{n=0}^{\infty}\frac{(3n)!}{3^nn!x^{3n+1}},\quad x\to\infty,   \eqno(10)
$$   
 $$
 \zeta=\frac23x^{3/2},\quad c_n=\frac{\Gamma(n+\frac16)
 \Gamma(n+\frac56)}{2^{n+1}n!\pi}=\frac{(6n-1)!!}{6^{3n}n!(2n-1)!!}.\eqno(10')
   $$

The relation (10) agrees with the relation
$$
{\rm Hi}(x)={\rm Bi}(x)-{\rm Gi}(x) \eqno(11)
$$
for the functions
$$
{\rm Bi}(x)=\int_0^{\infty}dt\{\exp(xt-\frac{t^3}3)
+\sin(xt+\frac{t^3}3)\},
                                                                   \eqno(12)
$$
$$
{\rm Gi}(x)=\int_0^{\infty}dt\sin(xt+\frac{t^3}3),                 \eqno(13)
$$
for which the asymptotic expansions are known [4]
$$
{\rm Bi}(x)\sim\pi^{1/2}x^{-1/4}e^{\zeta}\sum_{n=0}^{\infty}\frac{c_n}{\zeta^n},
\quad\zeta=\frac23x^{3/2}, \eqno(14)
$$
$$
{\rm Gi}(x)\sim\sum_{n=0}^{\infty}\frac{(3n)!}{3^nn!x^{3n+1}},\quad x\to\infty.
                                                                 \eqno(15)
$$
(Assuming ${\rm Bi}(x)=\frac12[{\rm Bi}(x+i\epsilon)+{\rm Bi}(x-i\epsilon)]$,
 $\epsilon\to0$,
it is easy  to show that the asymptotic expansion for ${\rm Bi}(x)$,
 $x\to+\infty$,
consists of only one series (14), cf. Chap.\,3 Sections 3.4 and 3.5 in [6] and
p.\,52 in [5].)

The asymptotic expansions for ${\rm Ai}_1(x)$ can be obtained by integration of 
the asymptotic expansion for ${\rm Ai}(t)$, which for $t\to+\infty$ has the form 
[3,4]
$$
{\rm Ai}(t)\sim\frac12\pi^{1/2}t^{-1/4}e^{-\zeta}\sum_{n=0}^{\infty}(-1)^n
\frac{c_n}{\zeta^n},
\quad\zeta=\frac23t^{3/2},                                       \eqno(16)
$$
$c_n$ are given in ($10'$). Using the definition (6) for ${\rm Ai}_1(x)$,
 let us 
replace the integration variable $t$ by $\zeta$. Then integrating by parts
the leading term on the right-hand side of (16), we have
$$
\int_v^{\infty}dt\,e^{-\zeta}t^{-1/4}=\left(\frac23\right)^{1/2}\int_{\xi}
^{\infty}d\zeta\,
e^{-\zeta}\zeta^{-1/2}=
$$
$$
\left(\frac23\right)^{1/2}e^{-\xi}\xi^{-1/2}-
\frac12\left(\frac23\right)^{1/2}
\int_{\xi}^{\infty}d\zeta\,e^{-\zeta}\zeta^{-3/2},\quad \xi=\frac23v^{3/2}.
$$
We unite the last (integral) term with the integral of term with $c_1$
on the right-hand side of (16) and integrate by parts again. Repeating 
the process, we get
$$
{\rm Ai}_1(v)\sim\frac12\pi^{1/2}v^{-3/4}e^{-\xi}\sum_{n=0}^{\infty}(-1)^n
\frac{a_n}{\xi^n},
\quad\xi=\frac23v^{3/2},\quad v\to\infty.                \eqno(17)
$$
Here $a_n$ are connected with $c_n$ by the recurrence relation
$$
a_n=c_n+\left(n-\frac12\right)a_{n-1}=
\sum_{k=0}^n\frac{\Gamma(n+\frac12)}{\Gamma(k+\frac12)}c_k=
\frac{\Gamma(n+\frac12)}{2\pi^{3/2}}\sum_{k=0}^n2^k
{\rm B}\left(k+\frac16,k+\frac56\right),  
$$
 $$
 a_0=1,\quad a_1=\frac{41}{2^3\cdot3^2},\quad a_2=\frac{9241}{2^7\cdot3^4},\quad
 a_3=\frac{5^2\cdot203009}{2^{10}\cdot3^7},\, \ldots \,.         \eqno(18)
 $$   

 It can be shown now that the asymptotic series for ${\rm Ai}_1(x)$,
  ${\rm Hi}(x)$, 
 ${\rm Ai}(x)$, see (17), (10), (16) and the series for derivatives
  ${\rm Hi}'(x)$, ${\rm Ai}'(x)$,
$$
{\rm Hi}'(x)\sim\pi^{1/2}x^{1/4}e^{\zeta}\sum_{n=0}^{\infty}\frac{d_n}{\zeta^n}
+\sum_{n=0}^{\infty}\frac{(3n+1)!}{3^nn!x^{3n+2}},   \eqno(19)
$$   
 $$
{\rm Ai}'(x)\sim-\frac12\pi^{1/2}x^{1/4}e^{-\zeta}\sum_{n=0}^{\infty}(-1)^n
\frac{d_n}{\zeta^n},
\quad d_n=-\frac{6n+1}{6n-1}c_n,                            \eqno(20)
$$
satisfy the relation (5). (Each series of a complete asymptotic expansion 
corresponds to its own analytic function, so one can differentiate asymptotic
series, see p.\,21 in [4], and consequently the complete asymptotic expansion).
To do this, we use the equality
$$
\sum_{n=0}^{2s}(-1)^nc_nd_{2s-n}=0, \quad s\ge1,  \eqno(21)
$$ 
equivalent to
$$
 \sum_{k=0}^{2s}(-1)^k\frac{\Gamma(k+\frac16)\Gamma(k+\frac56)
 \Gamma(2s-k-\frac16)
 \Gamma(2s-k+\frac76)}{k!(2s-k)!}=0,         \eqno(21')
 $$
 which can be regarded as a result of substitution of asymptotic series for 
 ${\rm Ai}(x)$ and ${\rm Bi}(x)$ in the Wronskian
 $$
 {\rm Bi}'(x){\rm Ai}(x)-{\rm Bi}(x){\rm Ai}'(x)=\pi.          \eqno(22)
 $$

 Using (11) and (22), the relation (5) can be rewritten in the form
 $$
 {\rm Ai}_1(x)= {\rm Gi}'(x){\rm Ai}(x)-{\rm Gi}(x){\rm Ai}'(x).    \eqno(23)
 $$
This can easily be checked by differentiation and using the equations
$$
{\rm Gi}''(x)-x{\rm Gi}(x)=-1,\quad {\rm Ai}''(x)-x{\rm Ai}(x)=0.
$$

With the help of (23) and asymptotic series (16), (20), (15) for ${\rm Ai}(x)$,
${\rm Ai}'(x)$, ${\rm Gi}(x)$ and the series
$$
{\rm Gi}'(x)\sim-\sum_{n=0}^{\infty}\frac{(3n+1)!}{3^nn!x^{3n+2}},
\quad x\to\infty,
                                                                 \eqno(24)
$$
obtained by differentiation, it is easy to recover the asymptotic expansions
(17) for ${\rm Ai}_1(x)$. The agreement of the series for ${\rm Ai}_1(x)$,
 obtained with
 the help of (23) and by direct integration of series for  ${\rm Ai}(x)$,
  is a check
of correctness of series (15), (24) for ${\rm Gi}(x)$ and ${\rm Gi}'(x)$.
 Now we note that 
the leading term of the asymptotic series for ${\rm Gi}'(x)$ given in
 handbook [3],
$$
{\rm Gi}'(x)\sim \frac7{96}x^{-2},\quad x\to\infty,
$$
disagrees with (24) and its integration (integration of asymptotic series is
 always possible [4]) does not reproduce the leading term of series for
 ${\rm Gi}(x)$.

It is worth-while to get the asymptotic expansion for ${\rm Ai}_1(x)$ with the
 remainder term. Assuming $v>0$ and using ${\rm Ai}''(x)=x{\rm Ai}(x)$, we have
 $$
 {\rm Ai}_1(v)=\int_v^{\infty}x^{-1}d{\rm Ai}'(x)=\left.x^{-1}{\rm Ai}'(x)
 \right|_v^{\infty}+
\int_v^{\infty}x^{-2}{\rm Ai}'(x)dx=                           \eqno(25)
$$
 $$
-v^{-1}{\rm Ai}'(v)+ \int_v^{\infty}x^{-2}d{\rm Ai}(x)=
-v^{-1}{\rm Ai}'(v)-v^{-2}{\rm Ai}(v)+
2\int_v^{\infty}x^{-3}{\rm Ai}(x)dx.
$$
Continuing the integration by parts with the help of the relation
$$
\int_v^{\infty}x^{-n}{\rm Ai}(x)dx=
-v^{-n-1}{\rm Ai}'(v)-(n+1)v^{-n-2}{\rm Ai}(v)+
$$
$$
(n+1)(n+2)\int_v^{\infty}x^{-n-3}{\rm Ai}(x)dx,          \eqno(26)
$$
we easily find
$$
 {\rm Ai}_1(v)=-\sum_{k=0}^{n}[\frac{(3k)!}{(3k)!!!v^{3k+1}}{\rm Ai}'(v)+
 \frac{(3k+1)!}{(3k)!!!v^{3k+2}}{\rm Ai}(v)]+
 $$
 $$
 +\frac{(3n+2)!}{(3n)!!!}\int_v^{\infty}x^{-3n-3}{\rm Ai}(x)dx, \quad
 (3n)!!!=3\cdot6\cdot9\cdot\,\ldots\,\cdot3n=3^nn!.                \eqno(27)
 $$
 For fixed $n$ and $v\to\infty$ the remainder term is negligible compared 
 with the terms of the sum. For $v\gg1$ and $n\to\infty$ we obtain again 
 the asymptotic expansion in the form (23) with the series for ${\rm Gi}(x)$
  and ${\rm Gi}'(x)$ given in (15) and (24).

 For $v<0$ we proceed similarly 
 $$
 {\rm Ai}_1(v)=\int_{-\infty}^{\infty}dx{\rm Ai}(x)-
 \int_{-\infty}^vdx{\rm Ai}(x)=\pi-
 \int_{-\infty}^vx^{-1}d{\rm Ai}'(x)=
 $$
 $$
 \pi-v^{-1}{\rm Ai}'(v)-v^{-2}{\rm Ai}(v)
 -2\int_{-\infty}^vx^{-3}{\rm Ai}(x)dx.                       \eqno(28)
 $$
 Finally we get
$$
 {\rm Ai}_1(v)=\pi-\sum_{k=0}^{n}[\frac{(3k)!}{(3k)!!!v^{3k+1}}{\rm Ai}'(v)+
 \frac{(3k+1)!}{(3k)!!!v^{3k+2}}{\rm Ai}(v)]+
 $$
 $$
 \frac{(3n+2)!}{(3n)!!!}\int_{-\infty}^vx^{-3n-3}{\rm Ai}(x)dx,
  \quad v<0.                                                 \eqno(29)
 $$

 Assuming $v\ll-1$ and letting $n\to\infty$, we obtain the asymptotic
 expansion in the form of the right-hand side of (5) with the series (4)
 for ${\rm Hi}(v)$ and the series for ${\rm Hi}'(v)$, obtained from (4) by differentiation:
    $$
    {\rm Hi}'(x)\sim\sum_{n=0}^{\infty}\frac{(3n+1)!}{3^nn!x^{3n+2}},\quad
     x\to-\infty.                                                  \eqno(30)
   $$

   We note that the leading term of the series for ${\rm Hi}'(x)$, given in [3],
    $$
    {\rm Hi}'(x)\sim-\frac32 x^{-2},\quad  x\to-\infty,   
   $$
   disagrees with (30) and its integration does not provide the leading term
    of (4) for $ {\rm Hi}(x)$.

 The analog of (17) for $v\to-\infty$ is
$$
{\rm Ai}_1(v)=\pi-\frac{\pi^{1/2}}{(-v)^{3/4}}\sum_{k=0}^{\infty}(-1)^k
[\frac{a_{2k}}{\zeta^{2k}}\cos(\zeta+\frac{\pi}4)+
\frac{a_{2k+1}}{\zeta^{2k+1}}\sin(\zeta+\frac{\pi}4)],
$$
$$
\quad\zeta=\frac23(-v)^{3/2},\quad v\to-\infty.                       \eqno(31)
$$
Here $a_n$ are the same as in (18).

To get the asymptotic expansion for ${\rm Gi}(-x)$, $x\to\infty$, we use (11)
and the asymptotic expansion for ${\rm Bi}(x)$, see [3] or (35) below. Then we
 obtain
 $$
{\rm Gi}(-x)\sim\frac{\pi^{1/2}}{x^{1/4}}\sum_{n=0}^{\infty}(-1)^n
[\frac{c_{2n}}{\zeta^{2n}}\cos(\zeta+\frac{\pi}4)+
\frac{c_{2n+1}}{\zeta^{2n+1}}\sin(\zeta+\frac{\pi}4)]+
\sum_{n=0}^{\infty}\frac{(3n)!}{3^nn!(-x)^{3n+1}},
$$
$$
\quad\zeta=\frac23x^{3/2},\quad x\to\infty.                       \eqno(32)
$$
Differentiating it , we have
 $$
{\rm Gi}'(-x)\sim\pi^{1/2}x^{1/4}\sum_{n=0}^{\infty}(-1)^n
[d_{2n}\zeta^{-2n}\sin(\zeta+\frac{\pi}4)-
d_{2n+1}\zeta^{-2n-1}\cos(\zeta+\frac{\pi}4)]-
$$
$$
\sum_{n=0}^{\infty}\frac{(3n+1)!}{3^nn!(-x)^{3n+2}}.           \eqno(33)
$$
$c_n$ and $d_n$ are the same as in ($10'$) and (20).
\section{Asymptotic expansions for ${\rm Ai}^2(x)$ and ${\rm Bi}^2(x)$
 for $x\to-\infty$}
In this case [3]
$$
(-x)^{1/4}{\rm Ai}(x)\sim S\sin(\zeta+\frac{\pi}4))-C\cos(\zeta+\frac{\pi}4),
                                                                \eqno(34)
$$
$$
(-x)^{1/4}{\rm Bi}(x)\sim S\cos(\zeta+\frac{\pi}4)+C\sin(\zeta+\frac{\pi}4),
\quad   \zeta=\frac23(-x)^{3/2},\quad x\to-\infty,  \eqno(35)
$$
where
$$
S=\sqrt{\pi}\sum_{k=0}^{\infty}(-1)^kc_{2k}\zeta^{-2k}, \quad 
C=\sqrt{\pi}\sum_{k=0}^{\infty}(-1)^kc_{2k+1}\zeta^{-2k-1}. \eqno(36)
$$
So the combinations
$$
\frac12[{\rm Ai}^2(x)\pm {\rm Bi}^2(x)]\equiv w_{1,2}(x) \eqno(37)
$$
are respectively the nonoscillatory and oscillatory parts of
 ${\rm Ai}^2(x)$.
Writing
$$
w_1(x)=\frac12[{\rm Ai}^2(x)+ {\rm Bi}^2(x)]\sim \frac{\pi}2\sum_{n=0}^{\infty}
\frac{e_n}{(-x)^{3n+1/2}}                                    \eqno(38)
$$
and substituting it in the equation 
$$
w'''(x)-4xw'(x)-2w(x)=0,                                   \eqno(39)
$$
satisfied by ${\rm Ai}^2(x)$ and ${\rm Bi}^2(x)$ [3], we get
$$
e_n=-\frac{(6n-1)(6n-3)(6n-5)}{2^5\cdot3n}e_{n-1}=
(-1)^n\frac{(6n-1)!!}{2^{5n}\cdot3^n\cdot n!}.                  \eqno(40)
$$
According to (34)-(36) $e_0=1$. Consequently
$$
w_1(x)=\frac12[{\rm Ai}^2(x)+ {\rm Bi}^2(x)]\sim 
\frac{\pi}2\sum_{n=0}^{\infty}
\frac{(-1)^n(6n-1)!!}{2^{5n}\cdot3^n\cdot n!(-x)^{3n+1/2}}.     \eqno(41) 
$$
Similarly for the oscillatory part of ${\rm Ai}^2(x)$ we write
$$
w_2(x)=\frac12[{\rm Ai}^2(x)-{\rm Bi}^2(x)]\sim 
$$
$$
\frac{\pi}2\sum_{n=0}^{\infty}
[\frac{g_{2n}}{(-x)^{3n+1/2}}\sin2\zeta+                 
\frac{g_{2n+1}}{(-x)^{3n+2}}\cos2\zeta],\quad
\zeta=\frac23(-x)^{3/2}.                                   \eqno(42)
$$
Substitution in (39) gives the recurrence relation
$$
g_n=\frac{(3n-5)(3n-3)(3n-1)}{2^5\cdot3\cdot n}g_{n-2}+
(-1)^n\frac{27n^2-27n+5}{2^3\cdot3\cdot n}g_{n-1}.              \eqno(43)
$$
The initial $g_0=1$ and  $g_1=-\frac5{2^3\cdot3}$ are easily obtainable from 
(34)-(36). The rest are determined from (43):
$$
g_2=-\frac{5\cdot41}{2^7\cdot3^2},\quad g_3=
\frac{5\cdot7\cdot11\cdot59}{2^{10}\cdot3^4},\quad g_4=
\frac{5\cdot7\cdot11\cdot12769}{2^{15}\cdot3^5}, \quad \ldots \;.   \eqno(44)
$$

\section{Stokes' phenomenon and the choice of saddle points}

In this Section using as an example the function $w(z)$, defined by the
 contour integral (45), we consider the Stokes' phenomenon, that is an abrupt
 appearance or disappearance of component series of a complete asymptotic
 expansion at certain phases of $z$. The complete asymptotic expansion 
 consists of several series; the number of these component series is different 
 in different sectors of the complex variable $z$. Each series corresponds 
 to a contribution from a single saddle point and determines its own function.
 So in different sectors of the complex plane $z$ the asymptotic expansion
 is determined by different number of saddle points. The problem of choosing of
 the saddle points, determining an asymptotic expansion, is connected with the
 topology of steepest decent lines and the disposition of ends of the 
 integration path in the integral representation for the function. For this
  reasons we draw the integration path along the steepest descent lines
  and will watch over its deformation with the change of $z$ in the complex
  plane. For definiteness we consider the solution of the Airy equation
  $w''=zw$ given by the integral 
  $$
  w(z)=i\int_Cdt\,e^{-i(zt+\frac{t^3}3)},  \eqno(45)
  $$
  where the contour $C$ goes from infinity in the sector $\pi/3<{\rm ph}\:
   t<2\pi/3$
  to infinity in the sector $-\pi/3<{\rm ph}\: t<0$ (sectors 1 and 3 in 
  fig.\,1-6).

  We note that $w(z)$ can be represented by the sum of two integrals over
  imaginary and real positive half-axes:
  $$
  w(z)=g(z)+f(z),\quad g(z)=i\int_{i\infty}^0dt\,e^{-i(zt+\frac{t^3}3)},\quad
  f(z)=i\int^{\infty}_0dt\,e^{-i(zt+\frac{t^3}3)}.                     \eqno(46)
  $$
  By substitution $t=i\tau$ the integral $g(z)$ is reduced to ${\rm Hi}(z)$.
   The integral $f(z)={\rm Gi}(z)+i{\rm Ai}(z)$. Then from (11) it follows
   $$
 w(z)={\rm Hi}(z)+{\rm Gi}(z)+i{\rm Ai}(z)={\rm Bi}(z)+i{\rm Ai}(z). \eqno(47)
 $$

 Introducing the modulus and phase of the parameter $z=|z|e^{i\varphi}$
 and the integration variable $t=|t|e^{i\theta}$, we write the exponent in (45)
 in the form
  $$
  -i(zt+\frac{t^3}3)=|t|[|z|\sin(\varphi+\theta)+\frac13|t|^2\sin3\theta]-
  i|t|[|z|\cos(\varphi+\theta)+\frac13|t|^2\cos3\theta].\eqno(48)
  $$
  The steepest decent lines are determined by the constancy of imaginary
  part of (48), that is by the condition
  $$
  -|t|[|z|\cos(\varphi+\theta)+\frac13|t|^2\cos3\theta]={\rm Const}.\eqno(49)
  $$
 The saddle points $t_{1,2}=|z|^{1/2}\exp[i(\varphi\mp\pi)/2]$. For the 
 steepest decent lines, passing over the saddle points $t_1, t_2$,
  the values of the constants on the right-hand side of (49) differ in sign:
 $$
 {\rm Const}=\mp\frac23|z|^{3/2}\sin\frac{3\varphi}2.   \eqno(50)
 $$
 Therefore the steepest decent lines from the passes $t_1, t_2$ meet and
 coincide below the lower pass only in exceptional cases, when the constant
  (50) becomes zero, i.e. when phase of the parameter $z$ is equal to
  $$
  \varphi=0,\; \pm\frac{2\pi}3,\; \pm\frac{4\pi}3,\; \ldots\; . \eqno(51)
  $$
  The lines of these exceptional values in the complex plane of the parameter
  $z$ are called the Stokes rays. Where go the ends of steepest descent lines
  for $|t|\to\infty$ ? From (49) it follows 
  $$
  |z|\cos(\varphi+\theta)+\frac13|t|^2\cos3\theta\to 0,\quad |t|\to\infty,
                                                                  \eqno(52)
  $$
  i.e.
  $$
  \theta\to\theta_{\infty}=\pm\frac{\pi}6(2k+1),\quad k=0, 1, 2,\; \ldots\; .
  $$
 Of these directions those ones, for which the real part of (48) goes to
 $-\infty$ for $|t|\to\infty$ (i.e. $\sin\theta<0$), just correspond to 
 descent and not to ascent. But 
 $$
 \sin[\pm\frac{\pi}2(2k+1)]=\pm(-1)^k
 $$
 is negative at odd $k$ for the upper sign and at even $k$ for the lower 
 sign. Hence
 $$
 \theta_{\infty}=-\frac{\pi}6,\;\frac{3\pi}6,\;-\frac{5\pi}6,\; \ldots\;.
                                                         \eqno(53)
 $$

 In fig.\,1-6 in the complex $t$-plane the locations of saddle points 
 $t_1, t_2$ and the behavior of steepest descent lines from the passes are
 qualitatively indicated for the specific values of phase of the parameter
 $z$, which changes in its own complex plane, see fig.\,7. The arrows mark off
 the integration path in (45) going from the sector $\pi/3<{\rm ph}\: t<2\pi/3$ into
 the sector $-\pi/3<{\rm ph}\: t<0$ along the steepest descent lines. When the
  parameter $z$ happens to be on a Stokes ray, the steepest descent lines,
  passing over two saddle points, meet one another at the lower saddle point.
  The integration path going over this saddle point either retains its 
  direction or abruptly changes it by $\pi/2$.

  In the latter case the integration path passes over both saddle points;
  the lower one (situated at the break) appears at the integration path 
  when the parameter $z$ crosses the Stokes ray. This is a manifestation
  of the Stokes'phenomenon: when $z$ crosses the Stokes ray the integration 
  path begins or stops to pass over the second (the lower) saddle point.

  At the absence of break on the integration contour it contains only one
   (lower) saddle point. In this case both before and after $z$ crosses
   the Stokes ray, the integration contour goes over only one (the lower)
   saddle point and Stokes phenomenon does not occur.

  In other words, the steepest descent lines, going over different saddle
  points, generally do not meet, but in one of the sectors their ends
  asymptotically approach each other. When the parameter $z$ approaches
  the Stokes ray, these ends come closer and closer at a greater length.
  For $z$ at the Stokes ray, they merge up to the lower saddle point; in
  this case the steepest descent line from the higher pass (active line)
  approaches the steepest descent line from the lower pass (passive line)
  at the angle of $\pi/2$. When $z$ goes beyond the Stokes ray, the active 
  line breaks away from the half of the passive line and goes to infinity      
  along the other half of the line. So, when $z$ crosses the Stokes ray,
  the passive line changes limply, but the active one drastically reverses
  the direction of its approach to the end of the passive line in order
  to approach the other end of that line. If the integration path for $z$
  near the Stokes ray contains the active line, then at $z$ crossing the 
  Stokes ray the passive line is included in (see fig.\,6, 1, 2) or excluded
  from (see fig.\,2, 3, 4) the integration path, bringing in it or out of it
  the lower saddle point. This is the Stokes' phenomenon.

 If the integration path for $z$ near the Stokes ray contains only the
 passive line of descent (see fig.\,4, 5, 6), then at $z$ crossing the
  Stokes ray, the active line (with its higher pass) is not contained 
  in the integration path, which, as before, goes only over the lower pass.

  So for $z$ outside the Stokes ray, the integration path contains one
   or two lines of steepest descent with the ends going to infinity. 
   Asymptotic expansion for the integral over each such line is obtained
   in usual manner by the saddle point method. For this reason, in order 
   to get the asymptotic expansion for the considered integral, it is 
   convenient to start from one of the expressions
   $$
   w(z)=w_{13}(z), \eqno(54)
   $$
   $$
   w(z)=w_{12}(z)+w_{23}(z),  \eqno(55)
   $$
   where $w_{ij}$ are defined by the same representation (45) as $w(z)$
   but with integration path $C_{ij}$  beginning in the $i-$ and ending 
   in the $j-$sector at infinity. These expressions hold for
   any $z$, but only one of them becomes suitable for the asymptotic
    expansion of the initial function in  that sector of the complex 
    $z-$plane for which each of the contours $C_{ij}$ can be drawn along
    the steepest descent line over the single saddle point.

   As seen from fig.\,1-6 for the sector $0<{\rm ph}\: t<2\pi/3$ the
    contours $C_{12}$,
   $C_{23}$ can be drawn along the steepest descent lines over the passes
   $t_2$, $t_{1}$. So in this sector the formula (55) is suitable. For the
   sector $2\pi/3<{\rm ph}\: t<2\pi$ or $-4\pi/3<{\rm ph}\: t<0$ this expression is unsuitable 
   as nether $C_{13}$, nor $C_{23}$ can be drawn along the steepest descent
   line over a single saddle point. However, for $z$ in this sector the
   contour $C_{13}$ can be drawn along the steepest descent line over the
 saddle point $t_1$. So here the expression (54) is suitable.

 If $z$ is on the Stokes ray and the integration contour, drawn along the
 steepest descent lines, does not suffer a break at the lower pass, then
 such a contour goes over the one, lower saddle point only, see fig.\,5, and
 for the asymptotic representation of integral $w(z)$ the formula (54) is
 suitable.

 If $z$ is on the Stokes ray and the integration contour, drawn along the
 steepest descent lines, suffers a break at the lower saddle point (and
 therefore passes the higher saddle point, see fig.\,1 or 3), then it is
 difficult to obtain the asymptotic expansion for the integral over this
 path. Yet this integral can be regarded as a limit of the half-sum of
 integrals with parameters $z_1$, $z_2$ lying on different sides of the
 Stokes ray and tending to $z$. Hence, for the asymptotic expansion 
 we can use the formula
 $$
 w(z)=\frac12[w_{12}+w_{23}+w_{13}]. \eqno(56)
 $$

For $z$ in the sector $-4\pi/3<{\rm ph}\: t<0$ and $|z|\gg1$, using (54)
 and the saddle point method, we represent $w(z)$ by the asymptotic series
$$
w(z) =w_{13}(z)=W_n^{(13)}(z)+R_n^{(13)}(z)\sim W_{\infty}^{(13)}(z), \eqno(57)
$$
where $W_n^{(13)}(z)$ is the sum of the first $n$ terms of the asymptotic
series $W_{\infty}^{(13)}(z)$, and $R_n^{(13)}(z)$ is the remainder term, small
in comparison with $W_n^{(13)}(z)\equiv 2S_n^{(2)}(z)$. The explicit terms
of $S_n^{(2)}(z)$ and $S_n^{(1)}(z)$, connected by the relation 
$S_n^{(2)}(z)=\pm iS_n^{(1)}(ze^{\pm2\pi i})$, are given in equations (68),
 (66).

 Similarly for $z$ in the sector $0<{\rm ph}\: t<2\pi/3$ and $|z|\gg1$ the
  function  $w(z)$ can be represented by two asymptotic series
 $$
 w(z)=w_{12}(z)+w_{23}(z)=                            \eqno(58)
 $$
 $$
W_n^{(12)}(z)+R_n^{(12)}(z)+W_n^{(23)}(z)+R_n^{(23)}(z)\sim W_{\infty}^{(12)}(z)
+W_{\infty}^{(23)}(z),
 $$
 forming together with (57) the complete asymptotic expansion.

 Near the Stokes ray ${\rm ph}\: z=0$ the series $W_{\infty}^{(12)}(z)$,
  representing
 the contribution from the higher pass, is dominant and the series 
 $W_{\infty}^{(23)}(z)$, representing the contribution from the lower pass,
 is recessive. On the ray ${\rm ph}\: z=\pi/3$  their role in the representation 
 for $w(z)$ is equal and near the Stokes ray ${\rm ph}\: z=2\pi/3$ the series
$W_{\infty}^{(23)}(z)$ becomes dominant and the series $W_{\infty}^{(12)}(z)$
recessive.

Near the Stokes ray ${\rm ph}\: z=0$ the series $W_{\infty}^{(13)}(z)$ in the 
representation (57) and the dominant series $W_{\infty}^{(12)}(z)$ in (58)
represent the contribution from the higher pass. Hence they differ only in
 their remainder terms, i.e. near the Stokes ray their $n-$th partial sums
 are equal:
 $$
 W_n^{(13)}(z)=W_n^{(12)}(z)\equiv 2S_n^{(2)}(z),          \eqno(59)
 $$
 but the remainder terms suffer a jump
 $$
R_n^{(13)}(z)-R_n^{(12)}(z)=w_{23}(z), \quad {\rm ph}\: z=0,    \eqno(60)
$$
equal to the analytic function $w_{23}(z)$. As this one is recessive, its
asymptotic representation and the remainder term does not suffer any jump at
this Stokes ray, cf. with the asymptotic behavior of $w_{13}(z)$ at the 
Stokes ray ${\rm ph}\: z=-2\pi/3$, eq. (57).

So on the Stokes ray ${\rm ph}\: z=0$ the asymptotic representation for
 $w(z)$ is
given by the $n-$th partial sum (59) of the dominant series and the remainder
 term
$$
R_n^{(13)}(z)=R_n^{(12)}(z)+w_{23}(z)=\frac12[
R_n^{(12)}(z)+R_n^{(13)}(z)+w_{23}(z)].                    \eqno(61)
$$
All three representations (61), which are the limits of representations (57)
 and (58) from each side of the Stokes ray and their half-sum, are equivalent.

Near the Stokes ray ${\rm ph}\: z=2\pi/3$, due to dominance of the higher pass, the
$n-$th partial sums $W_n^{(23)}(z)$ and $W_n^{(13)}(ze^{-2\pi i})$ are equal
(the branching of the sums at $z=0$ is taken into account):
$$
W_n^{(13)}(ze^{-2\pi i})=W_n^{(23)}(z)=2S_n^{(2)}(ze^{-2\pi i})=2iS_n^{(1)}(z),
                                                         \eqno(62)
$$
and the remainder terms suffer a jump at this ray
$$
R_n^{(13)}(ze^{-2\pi i})-R_n^{(23)}(z)=w_{12}(z),\quad {\rm ph}\: z
=\frac{2\pi}3.                                          \eqno(63)
$$
On the Stokes ray ${\rm ph}\: z=2\pi/3$ the asymptotic expansion for 
$w(z)$ is given
by the $n$-th partial sum (62) of the dominant series and the remainder
$$
R_n^{(13)}(ze^{-2\pi i})=R_n^{(23)}(z)+w_{12}(z)=
\frac12[
R_n^{(13)}(ze^{-2\pi i})+R_n^{(23)}(z)+w_{12}(z)].\eqno(64)
$$

Note that the functions $w_{ij}(z)$ can be turned one into another by rotation
of the integration path over the angle of $\pm2\pi/3$:
$$
w_{13}(z)=-e^{2\pi i/3}w_{23}(ze^{2\pi i/3}),\quad
w_{12}(z)=e^{-2\pi i/3}w_{23}(ze^{-2\pi i/3}).            \eqno(65)
$$
The function $w_{23}$ up to a factor coincides with the Airy function:
$w_{23}(z)=2i{\rm Ai}(z)$. As seen from fig.\,5, 6, 1, 2, 3 for $z$ in the sector 
$-2\pi/3<{\rm ph}\: z<2\pi/3$, the integration path $C_{23}$, drawn along the steepest
descent line, goes through one saddle point. So for $|z|\gg1$ the Airy function
is represented by one asymptotic series, the recessed one near the Stokes ray
${\rm ph}\: z=0$, cf. (57). By the saddle point method we get for this series
$$
{\rm Ai}(z)=S_n^{(1)}(z)+R_n(z),\quad |{\rm ph}\: z|\le\frac{2\pi}3,
$$
$$
S_n^{(1)}(z)=\frac{\pi^{1/2}}{2z^{1/4}}e^{-\zeta}\sum_{k=0}^{n-1}c_k
(-\zeta)^{-k},\quad \zeta=\frac23z^{3/2}.                            \eqno(66)
$$
For $|{\rm ph}\: z|\le2\pi/3$ the remainder term $R_n(z)$ is small in
 comparison with at least first few terms of $S_n^{(1)}(z)$. Outside this
  sector the value of the
representation (66) diminishes as $z$ approaches the negative half-axis
because $R_n(z)$ increases. In order to get better representation for 
the complementary sector $2\pi/3<{\rm ph}\: z\le4\pi/3$, we use the relation
$$
{\rm Ai}(z)=-e^{-{2\pi i}/3}{\rm Ai}(ze^{-{2\pi i}/3})-
e^{{2\pi i}/3} {\rm Ai}(ze^{-{4\pi i}/3}),      \eqno(67)
$$
following from (55) and (65). For variable $z$ in the complementary sector
the arguments of the Airy functions in the right-hand side of (67) does not
leave the main sector, where (66) holds. It is easy to verify that
$$
-e^{-{2\pi i}/3}{\rm Ai}(ze^{-{2\pi i}/3})=iS_n^{(2)}(z) -
e^{-{2\pi i}/3}R_n(ze^{-{2\pi i}/3}),
$$
 $$
S_n^{(2)}(z)=\frac{\pi^{1/2}}{2z^{1/4}}e^{\zeta}\sum_{k=0}^{n-1}c_k
\zeta^{-k}                                                    \eqno(68)
 $$
 and
 $$
-e^{{2\pi i}/3}{\rm Ai}(ze^{-{4\pi i}/3})=S_n^{(1)}(z) -
e^{{2\pi i}/3}R_n(ze^{-{4\pi i}/3}).                \eqno(69)
$$
Hence
$$
{\rm Ai}(z)=S_n^{(1)}(z)+S_n^{(1)}(ze^{-2\pi i})-e^{-{2\pi i}/3}
 R_n(ze^{-{2\pi i}/3})-                                    
 e^{{2\pi i}/3} R_n(ze^{-{4\pi i}/3}),
 $$
 $$
  2\pi/3\le{\rm ph}\: z\le4\pi/3.                           \eqno(70)
 $$
 As seen from here, the remainder, represented by the last two terms in the 
 right-hand side of (70), does not exceed the double maximum value of
 $R_n(z)$ in the main sector.

Similarly , we find
$$
w(z)=-2ie^{{2\pi i}/3} {\rm Ai}(ze^{{2\pi i}/3})=2S_n^{(2)}(z)+
2e^{{i\pi}/6}R_n(ze^{{2i\pi}/3}),\quad -4\pi/3\le {\rm ph}\: z<0,
                                                                    \eqno(71)
$$
 $$
 w(z)=2S_n^{(2)}(z)+2S_n^{(2)}(ze^{-2\pi i})+2iR_n(z)+2e^{-{i\pi}/6}
 R_n(ze^{-{2\pi i}/3}), \quad 0\le {\rm ph}\: z<2\pi/3.\eqno(72)
 $$  

 Asymptotic representations for ${\rm Bi}(z)$ follow from the given relations with
 account of equality ${\rm Bi}(z)=w(z)-i{\rm Ai}(z)$.

 As seen from (70) and (72), the asymptotic representation in the complementary
 sector is the sum of two analytical  continuations of the asymptotic series
  from the main sector, continuations corresponding to the right and the
left detours of the branch point $z=0$. Thereby the asymptotic representation
does not depend on the position of branch cut.

\section {Asymptotic behavior of $\int_v^{\infty}dx{\rm Ai}(x)h(x,v)$ for 
$v\to-\infty$}
   
The probabilities of many processes with elementary particles in a constant
external field are reduced to the integral [1,\,2,\,7]
$$
\int_v^{\infty}dx{\rm Ai}(x)h(x,v). \eqno(73)
$$
 For the
processes taking place also in the absence of a field, the parameter $v$ is
negative and $v\to-\infty$ for the field tending to zero. Let as consider the
asymptotic behavior of the integral (73) for $v\to-\infty$. It is assumed that 
the function $h(x,v)$ has no singularity in the integration interval and can be
expanded at $x=v$ in the series 
$$
h(x,v)=\sum_{k=-1}^{\infty}h_k\:(x-v)^{k/2}=
\sum_{k=-1}^{\infty}f_k\:\left(\frac{x-v}{-v}\right)^{k/2},   \eqno(74)
$$
in which the coefficients $f_k$ weakly depend on $v$,
$$
f_k\equiv h_k\cdot(-v)^{k/2}\sim f_0.                           \eqno(75)
$$
In the following {\it we omit for brevity the second argument of the function}
$h(x,v)$. Then we can write 
$$
\int_v^{\infty}dx\,{\rm Ai}(x)h(x)=h(0){\rm Ai}_1(v)+
\int_v^{\infty}dx\,x{\rm Ai}(x)\tilde h(x),
$$
$$
\quad \tilde h(x)=\frac{h(x)-h(0)}x=\sum_{k=-1}^{\infty}\tilde 
h_k\:(x-v)^{k/2}=
\sum_{k=-1}^{\infty}\tilde f_k\:\left(\frac{x-v}{-v}\right)^{k/2}.   \eqno(76)
$$
The function $\tilde h(x)$ has the property of $h(x)$: it is finite at $x=0$
and behaves as $\tilde f_{-1}\:\left(\frac{x-v}{-v}\right)^{-1/2}$ near
$x=v$.

Introducing the function
$$
H(x)=\tilde h(x)-\sum_{k=-1}^2\tilde f_k\:\left(\frac{x-v}{-v}\right)^{k/2}=
\sum_{k=3}^{\infty}\tilde f_k\:\left(\frac{x-v}{-v}\right)^{k/2},   \eqno(77)
$$
vanishing at $x=v$ not weaker than $(x-v)^{3/2}$, and using equation
$x{\rm Ai}(x)={\rm Ai}''(x)$, let us transform the second term in (76) by 
integration by parts:
$$
\int_v^{\infty}dx\,x{\rm Ai}(x)\tilde h(x)=
\tilde f_{-1}\sqrt{-v}\frac{d^2}{dv^2}
[2^{2/3}{\rm Ai}^2(t)]-\tilde f_0{\rm Ai}'(v)-
$$
$$
\frac{\tilde f_1}{2\sqrt{-v}}
\frac d{dv}[2^{2/3}{\rm Ai}^2(t)]-\tilde f_2\frac{{\rm Ai}(v)}v+
\int_v^{\infty}dx\,{\rm Ai}(x)H''(x), \quad t=2^{-2/3}v.    \eqno(78)
$$
Here we have used the equation [1]
$$
\int_v^{\infty}\frac{dx\,{\rm Ai}(x)}{\sqrt{x-v}}=
\int_v^{\infty}\frac{dx\,{\rm Ai}(x+v)}{\sqrt x}=2^{2/3}{\rm Ai}^2(t),
\quad t=2^{-2/3}v.                                     \eqno(79)
$$
It can be shown that the coefficients $\tilde f_k$ are connected 
with $f_k$ in (74) by formulae
$$
\tilde f_{2k-1}=\frac1v\sum_{s=0}^kf_{2s-1},\quad 
\tilde f_{2k}=\frac1v\sum_{s=0}^kf_{2s}-\frac1vh(0). \eqno(80)
$$
Using  (80), we obtain from (76) and (78)
$$
\int_v^{\infty}dx\,{\rm Ai}(x)h(x)=
h(0)[\pi-2\int_{-\infty}^v\frac{dx}{x^3}{\rm Ai}(x)]+
f_{-1}[\frac{{\rm Ai}(t){\rm Ai}'(t)}{(-v)^{3/2}}-\frac{t{\rm Ai}^2(t)+
{\rm Ai}'{}^2(t)}{\sqrt{-t}}]-
$$
$$
f_0[\frac{{\rm Ai}(v)}{v^2}+\frac{{\rm Ai}'(v)}{v}]+
f_1\frac{{\rm Ai}(t){\rm Ai}'(t)}{(-v)^{3/2}}-
f_2\frac{{\rm Ai}(v)}{v^2}+\int_v^{\infty}dx\,{\rm Ai}(x)h_1(x),
\quad t=2^{-2/3}v.   \eqno(81)
$$
Here $h_1(x)\equiv H''(x)$ and in the first term the equation (28) was used.
 The function
$h_1(x)$ near $x=v$ and $x=0$ has the properties of $h(x)$, but is $(-v)^3$ 
times less than the last one:
$$
h_1(x)\sim(-v)^{-3}h(x).  \eqno(82)
$$
Applying to the last integral on the right-hand side of (81) the same procedure
as to the original one and repeating this process, we obtain finally the
 asymptotic representation 
$$
\int_v^{\infty}dx\,{\rm Ai}(x)h(x)=
\sum_{k=0}^{\infty}\{h_k(0)[\pi-2\int_{-\infty}^v\frac{dx}{x^3}{\rm Ai}(x)]+
f_{k,-1}[\frac{{\rm Ai}(t){\rm Ai}'(t)}{(-v)^{3/2}}-\frac{t{\rm Ai}^2(t)+
{\rm Ai}'{}^2(t)}{\sqrt{-t}}]-
$$
$$
f_{k,0}[\frac{{\rm Ai}(v)}{v^2}+\frac{{\rm Ai}'(v)}{v}]+
f_{k,1}\frac{{\rm Ai}(t){\rm Ai}'(t)}
{(-v)^{3/2}}-
f_{k,2}\frac{{\rm Ai}(v)}{v^2}\},\quad t=2^{-2/3}v.   \eqno(83)
$$
Here, similarly to (74) we have used the notation 
$$
h_k(x,v)\equiv h_k(x)=\sum_{m=-1}^{\infty}h_{k,m}(x-v)^{m/2}=
\sum_{m=-1}^{\infty}f_{k,m}\:\left(\frac{x-v}{-v}\right)^{m/2};   \eqno(83')
$$  
$h_k(x)$ is obtained from $h_{k-1}(x)$ in the same manner as $h_1(x)$ 
from  $h_0(x)\equiv h(x)\equiv h(x,v)$. It is easy to obtain a connection 
of the coefficients $f_{k,m}$ with the coefficients of the previous function:
$$
f_{k,2m-1}=\frac{(m+\frac12)(m+\frac32)}{v^3}\sum_{s=0}^{m+2}f_{k-1,2s-1},  
$$
$$
f_{k,2m}=\frac{(m+1)(m+2)}{v^3}[\sum_{s=0}^{m+2}f_{k-1,2s}-h_{k-1}(0)].
                                                                    \eqno(84)
$$
With the help of these equations the odd coefficients $f_{k,2m-1}$ can be 
expressed via odd coefficients $f_{2s-1}$ of the initial function, and the
even coefficients $f_{k,2m}$ - via even coefficients $f_{2s}$ of function
 $h(x)$ and via values of functions $h_n(x)$, $n<k$, at $x=0$. 

 As the connection of $f_{k,2m}$ with  $f_{2s}$  and $h_n(0)$ is linear, let 
 us first find the dependence of $f_{k,2m}$ on $h_n(0)$ putting at first 
$f_{2s}=0$. So, using the second equation in (84) first for $k=1$, then
for $k=2$, we express $f_{2,2m}$ through $h(0)$ and $h_1(0)$. Continuing 
this process, we find 
$$
f_{k,2m}=-\frac{(m+2)!}{m!v^3}\{h_{k-1}(0)+
\sum_{n=0}^{m+2}\frac{(n+2)!h_{k-2}(0)}{n!v^3}+
\sum_{n=0}^{m+2}\frac{(n+2)!}{n!}
\sum_{l=0}^{n+2}\frac{(l+2)!h_{k-3}(0)}{l!v^6}+
\cdots 
$$
$$
+\sum_{n_{k-1}=0}^{m+2}\frac{(n_{k-1}+2)!}{n_{k-1}!}\cdots
\sum_{n_2=0}^{n_3+2}\frac{(n_2+2)!}{n_2!}
\sum_{n_1=0}^{n_2+2}\frac{(n_1+2)!h_0(0)}
{n_1!v^{3k-3}}\}=                                             
-\frac1{v^3}\sum_{s=0}^{k-1}Q_{m,s}\frac{h_{k-1-s}(0)}{v^{3s}},     \eqno(85)
$$
$$
Q_{m,s}=\frac{(m+2)!}{m!}\sum_{n_s=0}^{m+2}\frac{(n_s+2)!}{n_s!}\cdots
\sum_{n_2=0}^{n_3+2}\frac{(n_2+2)!}{n_2!}\sum_{n_1=0}^{n_2+2}
\frac{(n_1+2)!}{n_1!}=\frac{(m+2+3s)!}{(3s)!!!m!}. \eqno(86)
$$
The final formula for $Q_{m,s}$ is easily proved by the use of known formula
$$
\sum_{n=0}^{m}\frac{(n+p)!}{n!}=\frac{(m+p+1)!}{m!(p+1)}   \eqno(87)
$$
and the induction method. By definition, $(3n)!!!=
3\cdot6\cdot9\cdot\ldots\cdot(3n) $,
$0!!!=1$. Thus
$$
\sum_{k=0}^{\infty}f_{k,2m}=-\frac1{v^3m!}\sum_{s=0}^{\infty}\frac
{(3s+m+2)!}{(3s)!!!v^{3s}}\sum_{n=0}^{\infty}h_n(0).              \eqno(88)
$$
Then the combination of terms with even coefficients $f_{k,2m}$ in formula
(83) transforms into the following expression
$$
-\sum_{k=0}^{\infty}
f_{k,0}[\frac{{\rm Ai}(v)}{v^2}+\frac{{\rm Ai}'(v)}{v}]-\sum_{k=0}^{\infty}
f_{k,2}\frac{{\rm Ai}(v)}{v^2}=
$$
$$
=\sum_{n=0}^{\infty}h_n(0)\{\frac{{\rm Ai}'(v)}{v^4}
\sum_{s=0}^{\infty}\frac{(3s+2)!}{(3s)!!!v^{3s}}+\frac{{\rm Ai}(v)}{v^5}
\sum_{s=0}^{\infty}\frac{(3s+4)!}{(3s+3)!!!v^{3s}}\}.   \eqno(89)
$$
Using the relation ${\rm Ai}(x)=x{\rm Ai}''(x)$ and twice integrating by parts,
 we find
$$
 \int_{-\infty}^v\frac{dx}{x^n}{\rm Ai}(x)=\frac{{\rm Ai}'(x)}{v^{n+1}}+
 (n+1)\frac{{\rm Ai}(x)}{v^{n+1}}+(n+1)(n+2)
 \int_{-\infty}^v\frac{dx}{x^{n+3}}{\rm Ai}(x).
$$ 
Repeatedly using this equation, we obtain 
$$
2\int_{-\infty}^v\frac{dx}{x^3}{\rm Ai}(x)=\frac{{\rm Ai}'(v)}{v^4}
\sum_{s=0}^n\frac{(3s+2)!}{(3s)!!!v^{3s}}+\frac{{\rm Ai}(v)}{v^5}
\sum_{s=0}^n\frac{(3s+4)!}{(3s+3)!!!v^{3s}}+
\frac{(3n+5)!}{(3n+3)!!!}\int_{-\infty}^v\frac{dx}{x^{3n+6}}{\rm Ai}(x).
                                                                  \eqno(90)
$$
Thus the expression in braces in (89) is the asymptotic representation of 
integral (90). So for the  case $f_{2s}=0$, i.e. when only odd $k$ are present 
in (74), the expression (83) becomes simpler $(t=2^{-2/3}v$):
$$
\int_v^{\infty}dx\,{\rm Ai}(x)h(x)=
\sum_{k=0}^{\infty}\{\pi h_k(0)
-\frac{t{\rm Ai}^2(t)+{\rm Ai}'{}^2(t)}{\sqrt{-t}}f_{k,-1}+
\frac{{\rm Ai}(t){\rm Ai}'(t)}{(-v)^{3/2}}
(f_{k,-1}+f_{k,1})\}.                   \eqno(91)
$$
From the expression 
$$
h_{k+1}(x)=[\frac{h_k(x)-h_k(0)}{x}]''-\frac3{4v^3}
f_{k,-1}\left(\frac{x-v}{-v}\right)^{-5/2}+\frac1{4v^3}(f_{k,-1}+f_{k,1})
\left(\frac{x-v}{-v}\right)^{-3/2},                            \eqno(92)
$$
connecting the function $h_{k+1}(x)$ with the previous one, it follows
$$
h_{k+1}(0)=\frac13h_k^{(3)}(0)-\frac3{4v^3}f_{k,-1}+\frac1{4v^3}(f_{k,-1}+
f_{k,1}).                                                  \eqno(93)
$$
Substituting $k+1$ by $k$ in (92) and differentiating three times the obtained
$h_k(x)$, we find for $x=0$:
$$
h_k^{(3)}(0)=\frac16h_{k-1}^{(6)}(0)-\frac3{4v^3}f_{k-1,-1}
\cdot\frac52\cdot\frac72\cdot\frac92+\frac{f_{k-1,-1}+f_{k-1,1}}{4v^6}
\cdot\frac32\cdot\frac52\cdot\frac72.                       \eqno(94)
$$
Continuing to use formula (92) in such a way, it can be shown that
$$ 
h_{k+1}(0)=\frac1{(3k+3)!!!}h^{(3k+3)}(0)+
$$
$$
\sum_{n=0}^k\frac1{(3n)!!!}[-\frac3
{4v^{3n+3}}f_{k-n,-1}\left(\frac52\right)_{3n}+
\frac{f_{k-n,-1}+f_{k-n,1}}{4v^{3n+3}}
\left(\frac32\right)_{3n}],               \eqno(95)
$$
where the products of half-intiger numbers are written in terms of Pochhammer's
symbol
$$
(a)_n=a(a+1)(a+2)\cdots(a+n-1)=\frac{\Gamma(a+n)}{\Gamma(a)}.
$$
Then
$$
 \sum_{k=0}^{\infty}h_k(0)=\sum_{k=0}^{\infty}\{\frac{h^{(3k)}(0)}{(3k)!!!}-
 \frac3{4v^3}f_{k,-1}\cdot\phi_{5/2}(v)+\frac{f_{k,-1}+f_{k,1}}{4v^3}
 \phi_{3/2}(v)\},          \eqno(96)
 $$
 $$
\phi_a(v)=\sum_{n=0}^{\infty}\frac{\Gamma(a+3n)}{\Gamma(a)(3n)!!!v^{3n}}.
                                                \eqno(97)
 $$
 Thus the asymptotic representation of the initial integral takes the form
 $$
 \frac1{\pi}\int_v^{\infty}dx\,{\rm Ai}(x)h(x)=
\sum_{k=0}^{\infty}\{\frac{h^{(3k)}(0)}{(3k)!!!}-
[\frac{t{\rm Ai}^2(t)+{\rm Ai}'{}^2(t)}{\pi\sqrt{-t}}+
\frac3{4v^3}\phi_{5/2}(v)]f_{k,-1}+
$$
$$
 [\frac{{\rm Ai}(t){\rm Ai}'(t)}{\pi(-v)^{3/2}}+\frac{\phi_{3/2}(v)}{4v^3}]
(f_{k,-1}+f_{k,1})\},\quad  t=2^{-2/3}v,\quad (f_{2s}=0).                          \eqno(98)
$$
It only remains to express the coefficients $f_{k,\pm1}$ via the 
odd coefficients
$f_{2s-1}$ of function $h(x)$. It can be done by repeated employment of the
first formula in (84):
$$
f_{k,2m-1}=\frac{(m+\frac12)(m+\frac32)}{v^{3k}}
\sum_{n_{k-1}=0}^{m+2}(n_{k-1}+\frac12)(n_{k-1}+\frac32)\cdots
$$
$$
\cdots\sum_{n_2=0}^{n_3+2}(n_2+\frac12)(n_2+\frac32)
\sum_{n_1=0}^{n_2+2}(n_1+\frac12)(n_1+\frac32)\sum_{s=0}^{n_1+2}f_{2s-1}.
                                                                    \eqno(99)
 $$
 In this expression we reverse the order of summation first over $n_1$ and
 $s$, then over $n_2$ and $s$, and so on. Finally, we obtain      
 $$
 f_{k,2m-1}=\frac{(m+\frac12)(m+\frac32)}{v^{3k}}
 \sum_{s=0}^{m+2k}R_{2s-1}(k,m)f_{2s-1},                   \eqno(100)
 $$
 $$
R_{2s-1}(k,m)=\sum_{n_{k-1}=n_{k-1}(s)}^{m+2}(n_{k-1}+\frac12)(n_{k-1}+\frac32)
\cdots\sum_{n_2=n_2(s)}^{n_3+2}(n_2+\frac12)(n_2+\frac32)
\sum_{n_1=n_1(s)}^{n_2+2}(n_1+\frac12)(n_1+\frac32),  \eqno(101)
$$
$$
n_i(s)=\left\{\begin{array}{cc}
0,\quad 0\le s\le2i,\\
s-2i,\quad s\ge2i,  
\end{array}\right.\quad  R_{2s-1}(1,m)=1.
$$
The formula (98) is correct if the expansion of $h(x)\equiv h(x,v) $
 at $x=v$ contains only the odd coefficients $f_k$, see (74). Just such
  functions were considered in paper [2], 
where the representation (98) was obtained. For $f_{2s}\ne0$ it follows from 
the lower formula (84) that to the expression (85) for even coefficients
 $f_{k,2m}$, the sum 
 $$
 f_{k,2m}^+=\frac{(m+1)(m+2)}{v^{3k}}\sum_{s=0}^{m+2k}T_{2s}(k,m)f_{2s},
                                                              \eqno(102)
 $$
 analogous to the expression (100) for odd coefficients $f_{k,2m-1}$,
 is added. In this sum 
 $$
 T_{2s}(k,m)=\sum_{n_{k-1}=n_{k-1}(s)}^{m+2}(n_{k-1}+1)(n_{k-1}+2)
\cdots
$$
$$
\sum_{n_2=n_2(s)}^{n_3+2}(n_2+1)(n_2+2)
\sum_{n_1=n_1(s)}^{n_2+2}(n_1+1)(n_1+2),  \quad  T_{2s}(1,m)=1,  \eqno(103)
$$
and the $n_i(s)$ are the same as in (101). As a result in asymptotic
 representation of integral (83), the terms linear in ${\rm Ai}(v)$,
  ${\rm Ai}'(v)$ remain:
$$
 \frac1{\pi}\int_v^{\infty}dx\,{\rm Ai}(x)h(x)=
\sum_{k=0}^{\infty}\{\frac{h^{(3k)}(0)}{(3k)!!!}-
[\frac{t{\rm Ai}^2(t)+{\rm Ai}'{}^2(t)}{\pi\sqrt{-t}}+
\frac3{4v^3}\phi_{5/2}(v)]f_{k,-1}+
$$
$$
 [\frac{{\rm Ai}(t){\rm Ai}'(t)}{\pi(-v)^{3/2}}+\frac{\phi_{3/2}(v)}{4v^3}]
(f_{k,-1}+f_{k,1})-\frac{{\rm Ai}'(v)}{\pi v}f_{k,0}^+-
\frac{{\rm Ai}(v)}{\pi v^2}(f_{k,0}^++f_{k,2}^+)\},\quad  
t=2^{-2/3}v.                                               \eqno(104)
$$
We remind that $\phi_a(v)$, $a=3/2,\:5/2$, is given in (97) and also that 
$h(x)\equiv h(x,v)$ and $h^{(3k)}(0)=d^{3k}h(x,v)/(dx)^{3k}\left.\right|_{x=0}$
is a function of $v$.

\section{Comments to the formula (104)}

We note now that all terms of asymptotic representation (104), except 
$h^{(3k)}(0)$, are oscillatory functions of $v$. Indeed, the linear in  
${\rm Ai}(v)$, ${\rm Ai}'(v)$ terms for $v\ll-1$ are pure oscillatory according to 
(34). The quadratic in ${\rm Ai}(v)$, ${\rm Ai}'(v)$ terms for $v\ll-1$ 
contain both
oscillatory and nonoscillatory parts, but the last are canceled with 
nonoscillatory functions $\phi_a(v)$.

Really, as it was shown in Section 3, ${\rm Ai}^2(t)$ is the sum 
$$
{\rm Ai}^2(t)=w_1(t)+w_2(t)    \eqno(105)
$$
of nonoscillatory $w_1$ and oscillatory $w_2$ functions, the asymptotic 
series of which are given by formulas (41),(42). Differentiating (105), we
obtain
$$
{\rm Ai}(t){\rm Ai}'(t)=\frac12w_1'(t)+\frac12w_2'(t), \eqno(106)
$$
$$
t{\rm Ai}^2(t)+{\rm Ai}'{}^2(t)=\frac12w_1''(t)+\frac12w_2''(t), \eqno(107)
$$
It follows from (41) and (97) that
$$
\frac{w_1'(t)}{2\pi(-v)^{3/2}}=-\frac{\phi_{3/2}(v)}{4v^3},\quad
\frac{w_1''(t)}{2\pi\sqrt{-t}}=-\frac{3\phi_{5/2}(v)}{4v^3},  \eqno(108)
$$
so that combinations in square brackets in (104) are pure oscillatory
functions:
$$ 
\frac{{\rm Ai}(t){\rm Ai}'(t)}{\pi(-v)^{3/2}}+\frac{\phi_{3/2}(v)}{4v^3}=
\frac{w_2'(t)}{2\pi(-v)^{3/2}}=[-\frac1{2^3\cdot3\cdot v^3}
-\frac{5\cdot7\cdot89}
{2^8\cdot3^4\cdot v^6}+\cdots]\sin2\zeta+
$$
$$
[-\frac1{2(-v)^{3/2}}-\frac{5\cdot7}
{2^6\cdot3^3\cdot(-v)^{9/2}}+\frac{5\cdot7\cdot11\cdot13\cdot107}{2^{12}
\cdot3^5\cdot
(-v)^{15/2}}+\cdots]\cos2\zeta,                                 \eqno(109)
$$
$$
\frac{t{\rm Ai}^2(t)+{\rm Ai}'{}^2(t)}{\pi\sqrt{-t}}+\frac3{4v^3}\phi_{5/2}(v)=
\frac{w_2''(t)}{2\pi\sqrt{-t}}=
$$
$$
[-1-\frac1
{2^5\cdot3^2\cdot v^3}-\frac{5\cdot7\cdot3923}{2^{11}\cdot3^5\cdot v^6}
+\cdots]\sin2\zeta+
$$
$$
[-\frac1{2^3\cdot3\cdot(-v)^{3/2}}-\frac{5\cdot7\cdot19}
{2^7\cdot3^4\cdot(-v)^{9/2}}+\cdots]\cos2\zeta, \quad \zeta=\frac23(-t)^{3/2}=
\frac13(-v)^{3/2}.                                              \eqno(110)
$$

Let us further simplify the coefficients $R_{2s-1}(k,m)$ and $T_{2s}(k,m)$
in formulas for $f_{k,2s-1}$ and $f_{k,2s}^+$ at least for the lower values
of $k$, see (100-103). We begin with $T_{2s}(k,m)$. Using (87), we obtain for
the first sum in (103)
$$
\sum_{n_1=n_1(s)}^{n_2+2}(n_1+1)(n_1+2)=\sum_{n_1=0}^{n_2+2}\frac{(n_1+2)!}
{n_1!}-\theta\left(s-\frac52\right)\sum_{n_1=0}^{s-3}\frac{(n_1+2)!}{n_1!}=
$$
$$
=\frac{(n_2+5)!}{(n_2+2)!3}-\theta\left(s-\frac52\right)\frac{s!}{(s-3)!3}.
                                                                 \eqno(111)
$$
As $s$ takes only integer positive values, the step function $\theta$ may be
omitted because $(s-3)!$ turns into infinity at integer negative $s-3$.
Putting $n_2=m$, we have from (111) and (103)
$$
T_{2s}(2,m)
=\frac{(m+5)!}{(m+2)!3}-\frac{s!}{(s-3)!3}.                \eqno(112)
$$
For the second sum in (103) we obtain on account of (111)
$$ 
\sum_{n_2=n_2(s)}^{n_3+2}(n_2+1)(n_2+2)
\sum_{n_1=n_1(s)}^{n_2+2}(n_1+1)(n_1+2)=
$$
$$
=\frac{(n_3+8)!}{3\cdot6\cdot(n_3+2)!} 
-\frac{s!(n_3+5)!}{3^2\cdot(s-3)!(n_3+2)!}+
\frac{s!}{3\cdot6\cdot(s-6)!}.                                   \eqno(113)
$$
Hence, from (103) and (113) it follows for $k=3$ 
$$
T_{2s}(3,m)=\frac{(m+8)!}{3\cdot6\cdot(m+2)!} 
-\frac{s!(m+5)!}{3^2\cdot(s-3)!(m+2)!}+
\frac{s!}{3\cdot6\cdot(s-6)!}.                                   \eqno(114)
$$
Similarly,
$$
T_{2s}(4,m)=\frac{(m+11)!}{3\cdot6\cdot9\cdot(m+2)!} 
-\frac{s!(m+8)!}{3^2\cdot6\cdot(s-3)!(m+2)!}+
$$
$$
\frac{s!(m+5)!}{3^2\cdot6\cdot(s-6)!(m+2)!}
-\frac{s!}{3\cdot6\cdot9\cdot(s-9)!}, \eqno(115)
$$
and so on.

With the help of these expressions it is easy to find several first 
coefficients $f_{k,2m}^+$ in (102),
$$
f_{1,0}^+=2v^{-3}(f_0+f_2+f_4),
$$
$$
f_{1,2}^+=6v^{-3}(f_0+f_2+f_4+f_6),
$$
$$
f_{2,0}^+=4v^{-6}[10(f_0+f_2+f_4)+9f_6+6f_8],
$$
$$
f_{2,2}^+=12v^{-6}[20(f_0+f_2+f_4)+19f_6+16f_8+10f_{10}],
$$
$$
f_{3,0}^+=2^4\cdot5v^{-9}[2^2\cdot7(f_0+f_2+f_4)+3^3f_6+2^3\cdot3f_8
+2\cdot3^2f_{10}+3^2f_{12}],
$$
$$
f_{3,2}^+=2^4\cdot3\cdot5v^{-9}[2^2\cdot3\cdot7(f_0+f_2+f_4)+
2\cdot41f_6+2^2\cdot19f_8
+2^6f_{10}+3^2\cdot5f_{12}+3\cdot7f_{14}].      \eqno(116)
$$

To calculate $R_{2s-1}(k,m)$, we proceed similarly. Using the relation [8]
$$
\sum_{k=0}^n\frac{\Gamma(k+a)}{\Gamma(k+b)}=
\frac{\Gamma(n+a+1)}{(a-b+1)\Gamma(n+b)}-
\frac{\Gamma(a)}{(a-b+1)\Gamma(b-1)},                \eqno(117)
$$ 
we obtain for the first sum in (101)
$$
\sum_{n_1=n_1(s)}^{n_2+2}\left(n_1+\frac12\right)\left(n_1+\frac32\right)=
\frac{\Gamma(n_2+\frac{11}2)}{3\Gamma(n_2+\frac52)}+\frac18-
\theta\left(s-\frac52\right)[\frac{\Gamma(s+\frac12)}{3\Gamma(s-\frac52)}+
\frac18].                                                \eqno(118)
$$
Putting here $n_2=m$, we obtain $R_{2s-1}(2,m)$. To calculate $i$-th  sum, the
relation
$$
\sum_{n_i=n_i(s)}^{n_{i+1}+2}\frac{\Gamma(n_i+\frac52+3l)}{\Gamma(n_i+\frac12)}
=\frac1{3(l+1)}\{\frac{\Gamma(n_{i+1}+\frac52+3l+3)}{\Gamma(n_{i+1}+\frac52)}+
\frac{(6l+3)!!}{2^{3(l+1)}}-
$$
$$
\theta\left(s-\frac12-2i\right)[\frac{\Gamma(s-2i+\frac52+3l)}
{\Gamma(s-2i-\frac12)}+\frac{(6l+3)!!}{2^{3(l+1)}}]\}  \eqno(119)
$$
is useful.

For the second sum in (101), we find with the help of (118), (119)
$$
\sum_{n_2=n_2(s)}^{n_3+2}\frac{\Gamma(n_2+\frac52)}{\Gamma(n_2+\frac12)}
\sum_{n_1=n_1(s)}^{n_2+2}\frac{\Gamma(n_1+\frac52)}{\Gamma(n_1+\frac12)}=
\frac{\Gamma(n_3+\frac{17}2)}{3\cdot6\Gamma(n_3+\frac52)}+
\frac{\Gamma(n_3+\frac{11}2)}{3\Gamma(n_3+\frac52)}
\{\frac18-
$$
$$
\theta\left(s-\frac52\right)[\frac{\Gamma(s+\frac12)}
{3\Gamma(s-\frac52)}+
\frac18]\}+\frac{9!!}{3\cdot6\cdot2^6}+\frac1{2^6}-
\theta\left(s-\frac52\right)[\frac{\Gamma(s+\frac12)}
{3\cdot8\Gamma(s-\frac52)}+
\frac1{2^6}]+
$$
$$
\theta\left(s-\frac92\right)[\frac{\Gamma(s+\frac12)}
{3\cdot6\Gamma(s-\frac{11}2)}+
\frac{\Gamma(s+\frac12)}
{3\cdot8\Gamma(s-\frac52)}-\frac{9!!}{3\cdot6\cdot2^6}].
                                                            \eqno(120)
$$
At $n_3=m$ the right-hand side of (120) equals to $R_{2s-1}(3,m)$. The first
few coefficients $f_{k,2s-1}$ in (100) are
$$
f_{0,-1}=f_{-1},\quad f_{0,1}=f_{1},\quad f_{1,-1}=\frac3{4v^3}(f_{-1}+f_1+f_3),
\quad f_{1,1}=\frac{15}{4v^3}(f_{-1}+f_1+f_3+f_5),
$$
$$
f_{2,-1}=\frac3{16v^6}[53(f_{-1}+f_1+f_3)+50f_5+35f_7],
$$
$$
f_{2,1}=\frac{15}{16v^6}[116(f_{-1}+f_1+f_3)+113f_5+98f_7+63f_9].   \eqno(121)
$$

\section{Asymptotic behavior of integrals related to the integral (73)}

In applications the asymptotic behavior for $v\ll-1$ of integrals obtained
 from (73) by
the replacement of Airy function ${\rm Ai}(x)$  by the function ${\rm Ai}_1(x)$ or
${\rm Ai}'(x)$ are needed. In the first case, integrating by parts, we obtain
$$
\int_v^{\infty}dx\,{\rm Ai}_1(x)h(x)=
\int_v^{\infty}dx\,{\rm Ai}(x)g(x),\quad h(x)\equiv h(x,v),
\quad g(x)\equiv g(x,v),                                   \eqno(122)
$$                      
$$
g(x)=\int_v^xdx\,h(x)=2h_{-1}(x-v)^{1/2} +h_0(x-v)+\frac23h_1(x-v)^{3/2}
+\frac12h_2(x-v)^2+\cdots. \eqno(123)
$$
It is seen that $g(x)$ has the same structure as $h(x)$ in (74). Hence the
above consideration is applied to the right-hand side of (122).

In the second case we have
$$ 
\int_v^{\infty}dx\,{\rm Ai}'(x)[h_{-1}(x-v)^{-1/2}+\varphi(x)]=
$$
$$
=2h_{-1}{\rm Ai}(t){\rm Ai}'(t)
-h_0{\rm Ai}(v)-\int_v^{\infty}dx\,{\rm Ai}(x)\varphi'(x), \quad t=2^{-2/3}v,
                                                             \eqno(124)
$$
$$
h(x)=h_{-1}(x-v)^{-1/2}+h_0+h_1(x-v)^{1/2}+h_2(x-v)+\cdots=
h_{-1}(x-v)^{-1/2}+\varphi(x).                                  \eqno(125)
$$
The first term in the right-hand side of (124), arisen  from the first term 
on the right-hand side
of (125), is found by differentiation with respect to $v$ of the 
expression 
$$
\int_v^{\infty}dx\,{\rm Ai}(x)(x-v)^{-1/2}=
\int_0^{\infty}dt\,t^{-1/2}{\rm Ai}(t+v)=
2^{2/3}{\rm Ai}^2(2^{-2/3}v).                 \eqno(126)
$$
The function $\varphi'(x)$ according to (125) again has the structure of 
$h(x)$ in (74) and the preceding consideration is applicable to (124).

Note now that
$$
-\int_v^{\infty}dx\,{\rm Ai}(x)\varphi'(x)=
-\pi\sum_{n=0}^{\infty}\frac{\varphi^{3n+1}(0)}{(3n)!!!}+O.T., \eqno(127)
$$
where $O.T.$ are oscillatory terms. As
$$\frac{d^n}{dx^n}(x-v)^{-1/2}=\frac{(-1)^n(2n-1)!!}{2^n(x-v)^{n+1/2}},
$$
then 
$$
\varphi^{(3n+1)}(0)=
h^{(3n+1)}(0)+(-1)^n\frac{(6n+1)!!\,h_{-1}}{2^{3n+1}(-v)^{3n+3/2}}. \eqno(128)
$$
Hence,
$$
-\pi\sum_{n=0}^{\infty}\frac{\varphi^{(3n+1)}(0)}{(3n)!!!}=
-\pi\sum_{n=0}^{\infty}\frac{h^{(3n+1)}(0)}{(3n)!!!}-
\pi\sum_{n=0}^{\infty}\frac{(-1)^n(6n+1)!!\,h_{-1}}
{(3n)!!!2^{3n+1}(-v)^{3n+3/2}}.                                \eqno(129)
$$
On the other hand, for the nonoscillatory part of the  term 
 $2h_{-1}{\rm Ai}(t){\rm Ai}'(t)$ in the right-hand side of (124)
we have according to (41) and (106):
$$
2h_{-1}\frac12w_1'(t)\equiv\pi\sum_{n=0}^{\infty}\frac{(-1)^n(6n+1)!!\,h_{-1}}
{(3n)!!!2^{3n+1}(-v)^{3n+3/2}}, \quad t=2^{-2/3}v.     \eqno(130)
$$
So, the all nonoscillatory part (130) is cancelled by the second sum on the
right-hand side of (129). Consequently, the nonoscillatory part of (124) is
given by the first sum in the right -hand side of (129), and the 
oscillatory part can be found according to the previous consideration.

Note added to the electronic version of this paper.

The consideration of the recessive series switching on and off, carried by us in
the Lebedev Phys. Inst. Preprint N 253 (1985), and connected with the abrupt 
behavior of the steepest decent line over the lower pass, led us later to the
notion of the natural Stokes' line width [9-11].

The natural width of the Stokes line is defined by the dimension of the saddle
of the lower pass, i.e. by such change $\Delta \alpha$ of the parameter
$\alpha$  for which the steepest decent line from the higher pass essentially
 changes its  behavior near the lower pass, so that the difference of the
 phases
$$
{\rm Im}(f_1-f_2)=\Delta \omega\cdot\alpha +\cdots ,
\quad \Delta \omega=\omega_2(0)-
\omega_1(0),\quad \omega_{1,2}(\alpha)=-\partial f_{1,2}(\alpha)/\partial
\alpha,   \eqno(131)
$$
of the contributions $e^{f_1(\alpha)}$ and $e^{f_2(\alpha)}$  of the 
lower  and higher passes becomes perceptible quantity of the order of $1$
and satisfies the uncertainty relation:
$$
 \Delta \omega\cdot\Delta \alpha \gtrsim 1.\eqno(132) 
$$ 
\section*{References}
\begin{enumerate}
\item{}
Aspnes D.E., Phys.Rev., {\bf147}, 554 (1966).
\item{}
Ritus V.I., Zh. Eksp. Teor. Fiz. {\bf56}, 986 (1969)
[Sov. Phys. JETP  {\bf29}, 532 (1969)]; 
Trudy FIAN {\bf111}, 5 (1979); J. Sov. Laser
Res. {\bf6} (1985).
\item{}
Antosiewicz H.A., in {\sl Handbook of Mathematical Functions}, ed. by 
Abramowitz M. and Stegun I.A., Dover, New York (1965).
\item{}
Olver F.W.J., {\sl Asymptotics and Special Functions}, Acad. Press, London 
(1974).
\item{}
Dingle R.B., {\sl Asymptotic Expansions: Their Derivation and
 Interpretation},
Acad. Press, London and New York (1973).
\item{}
Heading J., {\sl An Introduction to Phase-Integral Method}, Mothuen, London,
John Wiley, New York (1962).
\item{}
Nikishov A.I., Ritus V.I., Trudy FIAN {\bf168}, 232 (1986) [{\sl Issues in
Intense-Field Quantum Electrodynamics}, p.\,300. Ed. by V.L. Ginzburg 
Nova Science Pub.,
Commack (1987)]. 
\item{}
Gradshteyn I.S., Ryzhik I.M., {\sl Tables of Integrals, Series and Products},
Acad. Press, New York (1967). 
\item{}
Nikishov A.I., Ritus V.I., Teor. i Mat. Fiz. {\bf92}, 24 (1992) [ English 
version: {\bf92}, 711 (1993)].
\item{}
Nikishov A.I., Ritus V.I., Third Intern. Workshop on Squeezed States
 and Uncertainty Relations. Proc. Workshop at the Univ. of Maryland
 Baltimore County, Baltimore, Maryland, August 10-13, 1993; pp. 307-314.
\item{}
Nikishov A.I., Ritus V.I., Zh. Eksp. Teor. Fiz. {\bf105}, 769 (1994)
[JETP {\bf78}, 411 (1994)].
\end{enumerate}
 \end{document}